# Vibrational Kinetics in Plasma as a Functional Problem: a Flux-Matching Approach

*Paola Diomede[1,*], Mauritius C. M. van de Sanden[1], Savino Longo[2]*

[1]*Center for Computational Energy Research, DIFFER - Dutch Institute for Fundamental Energy Research, De Zaale 20, 5612 AJ Eindhoven, the Netherlands*

[2]*Dipartimento di Chimica, Universita' degli Studi di Bari, via Orabona 4, 70126 Bari, Italy*

*Corresponding Author
E-mail: p.diomede@differ.nl
Tel: +31-(0)40-3334-924

**Abstract**

A new approach to calculate the vibrational distribution function of molecules in a medium providing energy for vibrational excitation is proposed and demonstrated. The approach is an improvement of solution methods based on the drift-diffusion Fokker-Planck (FP) equation for a double differentiable function representing the vibrational populations on a continuum internal energy scale. A self-consistent numerical solution avoids approximations used in previous analytical solutions. The dissociation flux, a key parameter in the FP equation, is fixed using the kinetics of molecular dissociation from near-continuum levels, so that the vibrational kinetics becomes a functional problem. The approach is demonstrated for the kinetics of asymmetric stretching of $CO_2$, showing that it represents an alternative, potentially much more efficient in computational terms, to the presently usual state-to-state approach which is based on the kinetics of the populations of individual levels, and gives complementary insight into the dissociation process.

**Introduction**

The population kinetics of vibrational states of molecules has found many applications in the past in fields like catalysis, laser chemistry, plasma (ionized gas) chemistry and chemistry of the interstellar medium. The main tool used in vibrational kinetics is the state-to-state (STS) finite rate method[1] based on the numerical solution of a Master Equation (ME). This last is actually a stiff system of non-linear ordinary differential equations, as many as the number of vibrational levels. When complex molecules, with thousands of vibrational levels, are involved, the pursuit for models of whole reactors including fluid dynamics and other aspects asks for alternative methods allowing a significant gain in computational speed.

The method of calculation proposed and used in this paper is based on the numerical integration of the drift-diffusion, or Fokker-Planck (FP) equation for the population of internal energy states, in this case vibrational energy. It is an improved version of the drift-diffusion approach used sparingly in the 70's and 80's as an alternative to the much more computational demanding ME approach.[2-6] The difference with respect to previous diffusion approaches is the use of a numerical solution which avoids the approximations of the perturbation techniques used in the past.[3,4] In a previous paper,[7] we have solved the Fokker-Planck (FP) equation numerically using a Monte Carlo (MC) technique based on the short time Green function of the same equation. In this paper, we use a much more effective method and analyse the role of the reactive flux $J$, which equals the dissociation rate per molecule at the steady state. We formulate a relaxation method based on the self-consistency of $J$ and the dissociation rate. This approach can be used to develop fast algorithms for the solution of vibrational kinetics and molecular dissociation problems in gases and gas discharges. To demonstrate our approach, we consider the very important test case of $CO_2$ dissociation in a low temperature, non-equilibrium plasma produced by an electric discharge. This has been an important system in the past for the development of high power $CO_2$ lasers.[8,9] In recent years much attention has been devoted in particular to low temperature $CO_2$ plasmas produced using renewable energy in the context of green chemistry: the key concept is the conversion of greenhouse $CO_2$ into more reactive species to be converted into new fuels or useful chemicals.[10] Even more recent is the idea of using $CO_2$ plasma reactors to produce oxygen from $CO_2$ in Mars colonization scenarios.[11]

## Computational method

The method proposed is based on the self-consistency of vibrational diffusion and molecular dissociation, i.e. the matching of the parameter *J* in the drift-diffusion problem with the dissociation rate along a vibrational coordinate with N+1 bound levels, from v = 0 to v = N, v being the vibrational quantum number. The dissociation rate is obtained from chemical kinetics based on reactions for the vibrational levels close to the continuum and the concept of the pseudo-level, i.e. a vibrational level in the continuum which actually represents the dissociated state of the molecules. To this aim, the first necessity is to establish an explicit form for the vibrational drift-diffusion, or FP, problem. We use the approach reported in the papers by Fridman and co-workers (e.g. Rusanov et al.[5]) and summarized in the books by the same author.[3,4,6] The same approach has been considered in our previous paper[7] and it is based on the drift-diffusion equation for the vibrational distribution *f*. The vibrational distribution in this approach is a doubly differentiable function, defined in such a way to match the number density of molecules in a vibrational level *n(v)* at the corresponding energy *ε(v)*, this last measured from the vibrational ground state, i.e. *f(ε(v)) = n(v)*. The drift-diffusion equation reads:

$$\frac{df}{d\varepsilon} = -\frac{d}{d\varepsilon}(af) + \frac{d}{d\varepsilon}(b+cf)\frac{df}{d\varepsilon}. \tag{1}$$

In this equation three transport coefficients *a*, *b* and *c*, which are functions of the vibrational energy $\varepsilon$, are introduced: *a* is the drift coefficient, *b* is the linear diffusion coefficient and *cf* is the non-linear diffusion coefficient. These coefficients can be calculated from the same kinetic data used in the STS ME by means of known formulas of stochastic processes:[12] details of such calculations can be found elsewhere[4,5,7,12] and are not reported here. Alternatively, the transport coefficients can be calculated directly from microscopic models of the energy transfer using the formulas reported in Brau:[13] this means that in perspective our method is not dependent on a previously formulated dataset for an STS model.

Eq. (1) can be conveniently rewritten in the form $dJ/d\varepsilon = 0$ where *J* is the total flux given by

$$J = af - (b+cf)\frac{df}{d\varepsilon}. \tag{2}$$

As a rigorous consequence of the FP equation, $J$, which is the dissociation rate per molecule (in s$^{-1}$), is constant along the whole vibrational energy axis. In this paper we keep the constant $J$ into the solution as a parameter to be fixed. The drift-diffusion equation can be rearranged as

$$\frac{d\ln f}{d\varepsilon} = \frac{a}{b+cf} - \frac{J}{(b+cf)f} \qquad (3)$$

which can be solved immediately by forward numerical integration from the boundary condition at $\varepsilon = 0$ although the unknown function $f(\varepsilon)$ appears in the right-hand side. Since $f(0)$ must equal $n(0)$ and $f(\varepsilon)$ is a Boltzmann distribution for small values of $\varepsilon$, the left boundary condition can be written $f(0) = n_{tot}/Z_{vib}$, where $n_{tot}$ is the total number density of molecules in any state, $Z_{vib}(T_v)$ is the vibrational partition function and $T_v$ is the low-energy vibrational temperature. The constant $J$ can be determined using its connection to the dissociation rate. In fact, dissociation occurs when molecules are excited beyond the dissociation threshold, by a pump-up process. Such a reaction has generally the form AB(N) + P → A + B + P where N is the last bound level below the continuum and P is a suitable reaction partner. The expression for the contribution to the dissociation rate of a suitable $p$-th process, e.g. $k_p n(N) n_p$, where $k_p$ is the related rate coefficient and $n_p$ the number density of the reaction partner, can be equated to $J n_{tot}$ where $n_{tot}$ is the total number density of AB molecules. Since $f(\varepsilon_{max}) \sim n(N)$, where $\varepsilon_{max}$ is the energy of the last (v = N) bound level, the relation between $J$ and $f(\varepsilon_{max})$ is found in the form

$$J = f(\varepsilon_{max}) \cdot \frac{1}{n_{tot}} \sum_p k_p n_p \qquad (4)$$

Eq. (4) of course describes a very simple model of depletion by dissociation. In more complex chemical networks, sometimes more appropriate, dissociation can occur from states of lower energy than the uppermost bound one. For such networks the right-hand side of Eq. (4) is an integral operator acting on $f$.

This equation is the closure of the vibrational transport problem, linking the mesoscopic description provided by $f$ to the microscopic dissociation rate. In this approach, the vibrational kinetics problem becomes a functional one: a global property of the sought function is related to a single value of the same solution through a chemically meaningful boundary condition. Although a similar problem has been formulated in the past,[3,4] the solution proposed was approximated: a zero-order solution $f_0$ was calculated assuming $J = 0$ and then a first order one was written in the form $f_0 I$ where $I$ is an integral expression providing an estimate of the dissociation term based on $f_0$ (Eq. 7.12 in Rusanov and Fridman[3] eq. 3-174 and 3-176 in Fridman[4]). This approach was the basis for a few published solutions, but of course modern computational techniques allow the direct solution of the functional problem.

The functional problem is well posed and can be solved after the values of the rate coefficients are fixed. Since $f$ appears in the expression of $J$, the solution must be determined by a relaxation method: Eq. (3) is solved for a constant $J$ assumed initially equal to zero. Then $J$ is determined from the calculated $f(\varepsilon_{max})$ and the solution procedure is repeated, until self-consistency is obtained. The method is very fast, since only very few iterations are found to be necessary in real cases.

**Results and discussion**

In the case of $CO_2$ molecules, three vibrational modes have to be accounted for: symmetric stretching, doubly degenerate bending and asymmetric stretching. In plasma conditions, the detailed discussion in Fridman[4] (see also Kozák et al.[14]) shows that the most important contribution to dissociation is given by vibrational excitation of the asymmetric stretching mode. Therefore, we will focus on the kinetics of this mode only. Usually 21 (22 including the ground state) levels are considered for this mode, up to the dissociation energy of 5.5 eV. Processes included in the model are summarized in Table 1. Here, following notation in Kozák et al.,[14] $VV_1$ indicates linear vibration to vibration energy exchanges with the v = 1 level, $VV_n$ non-linear vibration to vibration energy exchanges, $VT_a$, $VT_b$, $VT_c$ vibration to translation exchanges of asymmetric mode levels with a, b, c symmetric levels, VV' vibration to vibration exchanges of asymmetric mode levels with symmetric levels, these last considered as vibration to translation (VT) processes as also recommended by Kozák et al..[14] STS rate coefficients need to be interpolated as a function of a continuum vibrational energy to be used in the diffusion equation. Details of this process are in Diomede et al..[7] A difference here with respect to our previous work[7] is that the contribution to the diffusion coefficient of the VT processes ($b_{VT}$) has been included using detailed balance (Eq. (6) in Diomede et al.[7]), in agreement with the theory in Rusanov et al..[3] Mathematical expressions for the transport coefficients and analytical fits for the STS rate coefficients are reported in Table 2 and 3, respectively. Recent works suggest that this model should be completed by a set of detailed inter-mode vibrational energy transitions (see e.g. Armenise and Kustova[15]). However, in this paper we will use the kinetic model by Kozák et al.[14] to benchmark our approach as we did in in our previous paper.[7]

Dissociation occurs when molecules are vibrationally excited beyond their last bound level, by pump-up processes like, in this particular case, vibrational exchange $VV_1$ and $VV_n$ processes in Table 1. Here only one quantum processes are assumed, the short discussion following Eq. (4) applies otherwise.

Transport coefficients obtained by applying the equations in Table 2 are shown in Figure 1. Several conditions are reported which differ for the values of the vibrational temperature $T_v$. In particular, three values for $T_v$ (for the asymmetric stretching mode), realistic for plasma activation reactors have been considered, namely 0.19, 0.22 and 0.25 eV, while the gas temperature $T_g$ was held fixed to 300 K, in order to avoid double interpolation ($T_v$ and $T_g$) of reaction kinetics data. $T_v$ can be calculated in plasma conditions as discussed in Capitelli[1] but here it is considered as a parameter. A notable feature, already discussed in Diomede et al.,[7] is the change of sign of the contribution to the drift coefficient due to VV processes in correspondence to the energy of the Treanor minimum given by $x_e \varepsilon_{1,0} T_v / 2T_g$, where $x_e$ is the coefficient of anharmonicity, $\varepsilon_{1,0}$ is the energy difference between the first two vibrational levels. This change of sign is essential in producing a characteristic change of trend of the solution $f$ to be discussed later. For energies higher than this critical one, the drift values are strongly sensitive to the value of the temperature ratio $T_v/T_g$.

In our case, the most important dissociation channels are via $VV_1$ processes which lead to dissociation when a $CO_2(1)$ molecule reacts with a molecule in the last level of the series $CO_2(v)$ producing $CO_2(0)$ and $CO_2(N+1)$, N+1 indicating the vibrational pseudo-level,[12] and $VV_n$ processes which lead to dissociation when a $CO_2(N)$ molecule reacts with a molecule in the last level of the series $CO_2(v)$ producing $CO_2(N-1)$ and $CO_2(N+1)$. Therefore, by applying Eq.(4), the following expression for J is obtained:

$$J = k_{N,N+1}^{1,0} \frac{e^{-\frac{\varepsilon_{1,0}}{kT_v}}}{Z_{vib}} f(\varepsilon_{max}) + k_{N,N+1}^{N,N-1} f(\varepsilon_{max})^2 / n_{tot}; \qquad (5)$$

Here $k_{i',j'}^{i,j}$ is a VV rate coefficient where *(i,j)* and *(i',j')* address the initial and final vibrational state of the two molecules involved. *J*, being the flux of dissociating molecules, must be consistent with that calculated from finite rate kinetics applied to the levels close to continuum.

**Table 1.** Elementary Reactions Used in the Calculations.

| name | reaction | note |
|---|---|---|
| $VV_1$ | $CO_2(1) + CO_2(v) \rightarrow CO_2(0) + CO_2(v+1)$ | |
| $VV_n$ | $CO_2(v) + CO_2(v) \rightarrow CO_2(v-1) + CO_2(v+1)$ | |
| VT | $CO_2(v) + CO_2 \rightarrow CO_2(v-1) + CO_2$ | a |
| VV' | $CO_2(v) + CO_2 \rightarrow CO_2(v-1) + CO_2$ | b |

[a] Sum $VT_a + VT_b + VT_c$ in Kozák et al.[14]

[b] Sum $VV'_a + VV'_b$ in Kozák et al.[14]

**Table 2.** Mathematical Expressions for Transport Coefficients Reported in Fig. 1.

$$b_{VV_1}(\varepsilon) = \frac{1}{2} k_{VV_1}(\varepsilon) n_{tot} (\hbar\omega)^2 \frac{P_1}{Z_{vib}}, \quad P_1 = \frac{P(1)}{P(0)} = \exp\left(-\frac{\varepsilon_{1,0}}{T_v}\right)$$

$$b_{VV_n}(\varepsilon) = c_{VV_n}(\varepsilon) f(\varepsilon), \quad c_{VV_n}(\varepsilon) = \frac{1}{2} k_{VV_n}(\varepsilon)(\hbar\omega)^2$$

$$a_{VT}(\varepsilon) = -(\hbar\omega) k_{VT}(\varepsilon) n_M$$

$$a_{VV'}(\varepsilon) = -(\hbar\omega) k_{VV'}(\varepsilon) n_M$$

$$a_{VV}(\varepsilon) = -(b_{VV_1}(\varepsilon) + b_{VV_n}(\varepsilon)) \left(\frac{1}{T_v} - \frac{2x_e \varepsilon}{T_g \hbar\omega}\right)$$

$$b_{VT}(\varepsilon) = (k_{VT}(\varepsilon) + k_{VV'}(\varepsilon)) n_M (\hbar\omega)^2$$

**Table 3.** Analytical Fits for the STS Rate Coefficients (in m³s⁻¹ with $\varepsilon$ in eV) (from Kozák et al.,[14] for $T_g$ = 300 K).

| fit | $c_0$ | $c_1$ | $c_2$ | $c_3$ | $c_4$ | note |
|---|---|---|---|---|---|---|
| $k_{VV_1}(\varepsilon) = 10^{-6} 10^{\sum_{i=0}^{4} c_i \varepsilon^i}$ | -9.96058 | 0.51516 | -0.38343 | 0.06286 | -0.00407 | |
| $k_{VV_n}(\varepsilon) = 10^{-6} 10^{\sum_{i=0}^{4} c_i \varepsilon^i}$ | -10.23249 | 1.91896 | -0.69428 | 0.13232 | -0.00948 | |
| $k_{VT}(\varepsilon) = 10^{-6} \sum_{i=0}^{4} c_i \varepsilon^i$ | $1.09698 \times 10^{-15}$ | $8.98604 \times 10^{-15}$ | $-3.92225 \times 10^{-15}$ | $7.48785 \times 10^{-16}$ | $-4.36492 \times 10^{-17}$ | a |
| $k_{VV'}(\varepsilon) = 10^{-6} 10^{\sum_{i=0}^{4} c_i \varepsilon^i}$ | -14.61444 | 1.44670 | -0.45651 | 0.09481 | -0.00691 | b |

[a] Sum $VT_a + VT_b + VT_c$ in Kozák et al.[14]

[b] Sum $VV'_a + VV'_b$ in Kozák et al.[14]

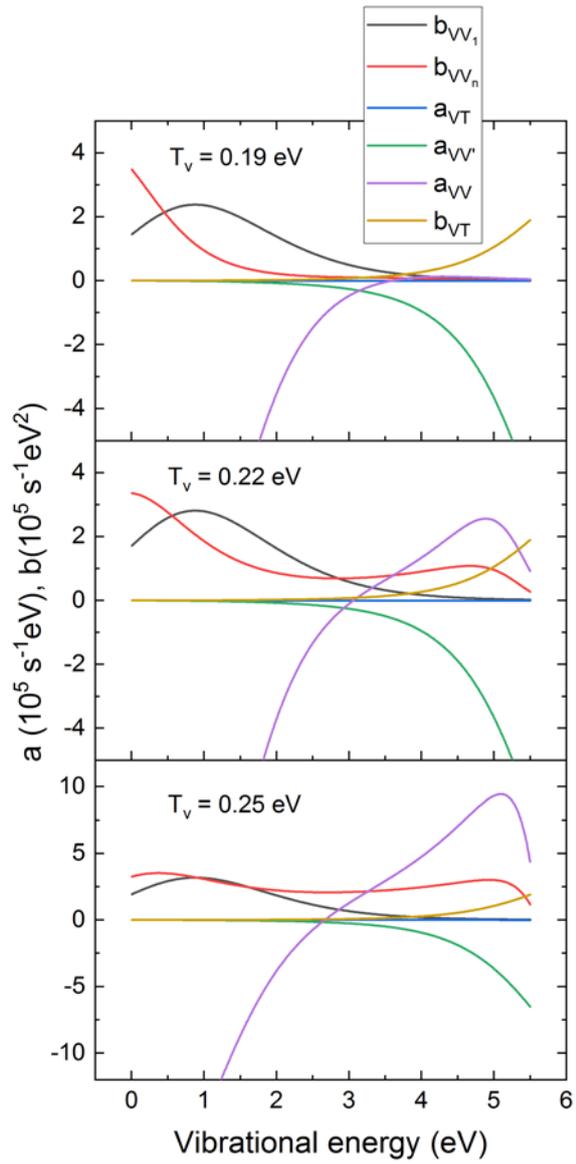

**Figure 1.** Coefficients *a* and *b* for different values of the vibrational temperature ($T_g$ = 300 K, $n_{tot}$ = 2.33 × $10^{23}$ $m^{-3}$) and flux-matching value of *J*.

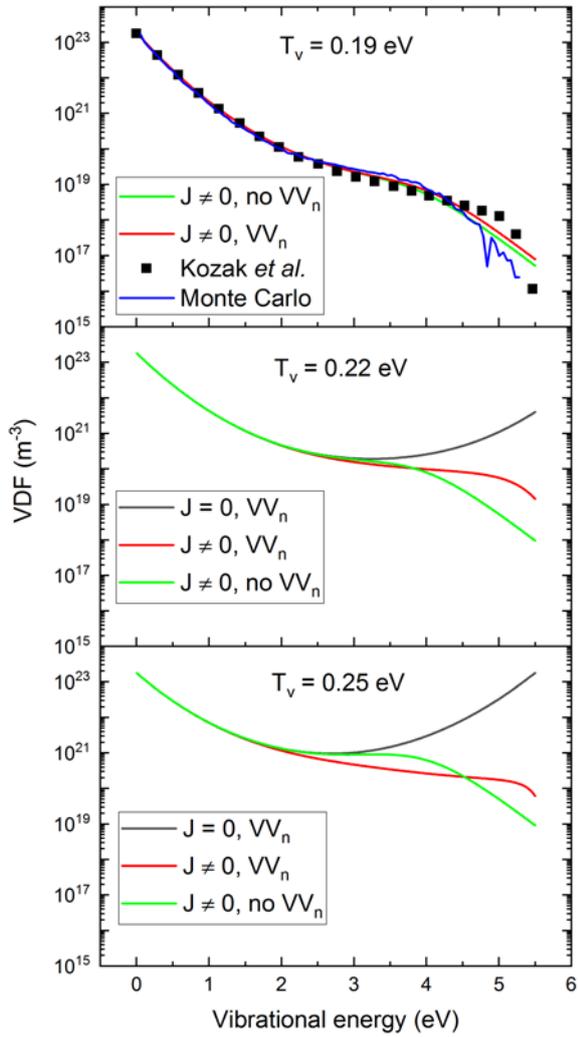

**Figure 2.** Vibrational distribution functions for different values of the vibrational temperature and null and non-null flux and including and excluding $VV_n$ processes (Same conditions as Figure 1). For $T_v$ = 0.19 eV, a comparison with STS results in Figure 7 in Kozák et al.[14] (for 30 W cm$^{-3}$ power density and 8 ms) and Monte Carlo simulations results is also shown. Results for $J$ = 0 and $T_v$ = 0.19 eV overlap with results for a non-null $J$ and without $VV_n$ processes, therefore they are not shown. Note that $J$ is different in the cases with and without $VV_n$ processes.

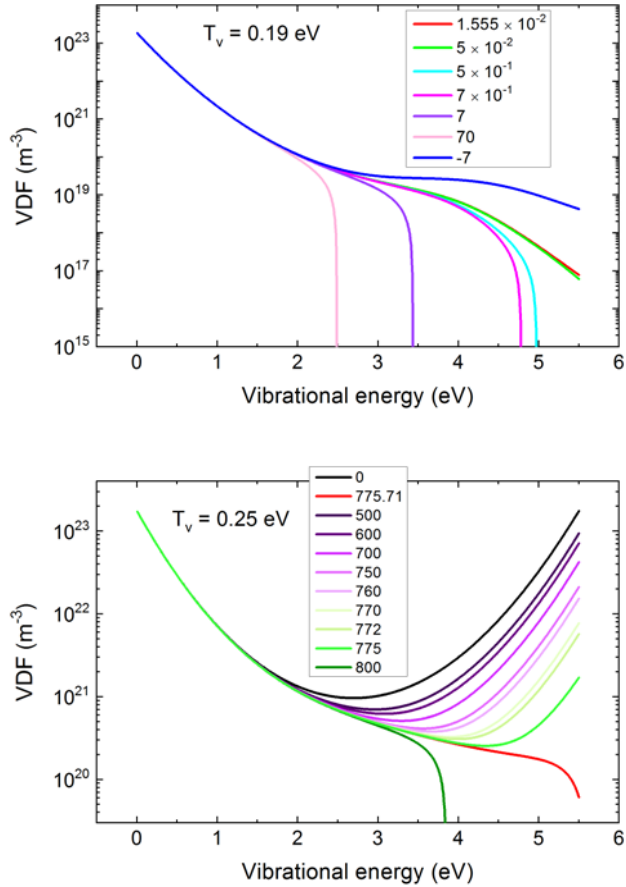

**Figure 3.** Vibrational distribution functions for different values of the flux $J$ (in $s^{-1}$) and 2 values of the vibrational temperature: 0.19 eV (top) and 0.25 eV (bottom), with flux-matching value of $J$ equal to $1.555 \times 10^{-2}$ $s^{-1}$ and 775.71 $s^{-1}$ respectively. Non-linear $VV_n$ processes are included.

In Figure 2 vibrational distribution functions (VDFs) corresponding to the same conditions in Figure 1, are shown. The non-linear (including $VV_n$) and linear equations are considered. The inclusion of $VV_n$ processes increases $J$ leading to a slight depression of the plateau and the depletion of the highest levels as discussed later. With regards to the boundary conditions, the flux-matched values of $J$ are considered. In the same Figure 2 we have also reported the results for the VDF obtained in recent calculations by Kozák et al.[14] in order to show how the diffusion approach compares to STS calculations. Due to an improved interpolation of kinetics data and a few differences in the calculation of transport coefficients (see above), our result here is not exactly matching the MC approach in Diomede et al..[7] In particular, we obtain here somewhat different behaviours of the distribution close to the dissociation threshold. Therefore, new MC calculations have been performed with the same set of data used here and results are reported in the same figure. It can be seen in this way that the flux-matching approach is in good agreement with the MC method. The residual difference very close to the dissociation threshold is due to the limited precision of our MC calculations, using a single time step, in the high energy region where the convective transport is fast. The comparison with STS is good, some deviations at high energy may be due to the fact that the calculations by Kozák et al.[14] include a few other processes involving both $CO_2$ and other species. Nevertheless, it is shown here that the diffusion approach, which is very fast in producing numerical results, represents a very good alternative to STS calculations for full reactors models where the computational cost is a critical issue. Results in Figure 2 also show that, under conditions of higher vibrational temperature, the appropriate consideration of the vibrational flux is a necessity. Results in Figure 2 also illustrate the role played by non-linearities introduced in the drift-diffusion equation by resonant processes: such processes play a small role (and may be neglected) under conditions of low $T_v$ and low $J$, while, when increasing $T_v$, they must be accounted for. Under such conditions the resonant process contributes to the dissociation significantly and should be included in the formulation of the boundary conditions. VT processes contribute to both diffusion and drift in the region of high energy where their rate becomes important. However they produce a sensible depletion of the high energy VDF only at the lowest vibrational temperature 0.19 eV (not shown).

In order to better illustrate the approach, calculations have been performed by assuming different values for $J$: results are reported in Figure 3. Two realistic values for $T_v$ are selected. What can be seen is not only the extreme importance of the selection of a correct value for $J$, but also the extreme sensitivity of the results to $J$. Furthermore, since a different value of $J$ amounts to a different relation between $f$ and its gradient $df/d\varepsilon$ at the boundary $\varepsilon = \varepsilon_{diss}$, this result confirms the findings in our previous paper[7] that boundary conditions play a fundamental role in continuum formulations of the vibrational kinetics. In agreement with the findings of our previous paper,[7] in Figure 3 (bottom) the assumption of a null flux $J$ produces a distribution very close to Treanor. A change in the VDF trend occurs, as previously anticipated, in correspondence to the minimum of this Treanor distribution. For the flux-matched solution this change leads to a plateau extending almost to the dissociation threshold: this is explained as the result of an essentially convective transport. Different curves are obtained by changing the reactive processes included in the high energy region. For example, the lowest curves corresponding to high $J$ require reactive channels to remove molecules at relative low energy. The curve with negative *J* requires a recombination channel, which populates the VDF close to the dissociation limit and is consistent with the negative flux: this last result also shows that our approach can be applied also in a recombination regime.

In a comparison with other methods of solutions, it must be remarked that the solution of the transport equation using the numerical integration of Eq. (2) has a very low computational cost, and the flux matching approach requires just a few iterations of the process. Furthermore, future implementations could be much less dependent on rate coefficients calculated for STS codes: it is in fact well known that many essential coefficients can be calculated reliably by using classical methods like molecular dynamics, and in the past Brau[13] has shown that the two coefficients *a* and *b* can be determined directly from such calculations, without artificially separating the results into discrete vibrational levels. Therefore, the approach presented in this paper is very promising for the development of complete models of plasma systems to be employed in concrete applications.

## Conclusions

In conclusion, an improved diffusion approach is developed and demonstrated for the kinetics of the asymmetric vibrational mode in $CO_2$ molecules. While the basic ideas of the drift-diffusion approach are well established in the literature, the approach is made of practical use by employing numerical techniques, to consider in full extent the effect of the reactive flux *J* and non-linear processes. The vibrational kinetics is therefore formulated as a functional problem based on the consistency of the VDF with the microscopic formulation of the reactive flux. In this way, a general, new technique is obtained which is a valid alternative to the presently widespread and much actively investigated STS approach and allows concretely to solve the vibrational problem with very low computational cost, as requested in multi-physical models of plasma reactors. In terms of added insight for chemical kinetics, this approach sets the vibrational kinetics into the field of transport processes, allowing to employ well developed concepts and mathematical techniques of the latter. In view of its simplicity and low computational cost, this approach may find application in any field where vibrational kinetics plays a role, thereby in catalysis, space chemistry, laser chemistry, plasma processes, and many others.

## Acknowledgements

This work is part of the Shell-NWO/FOM initiative 'Computational sciences for energy research' of Shell and Chemical Sciences, Earth and Life Sciences, Physical Sciences, FOM and STW, project number 14CSTT02, "Fast and accurate computational approaches to molecular dissociation in non-equilibrium plasmas: the case for $CO_2$ dissociation".

**TOC Graphic**

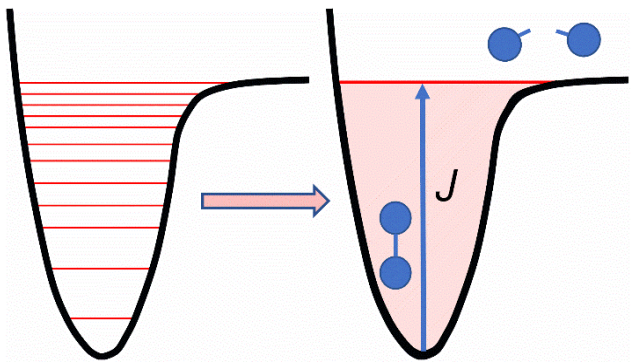